\documentstyle[12pt]{article}
\begin{document}
\baselineskip=14pt

\begin{center}
{\large MULTI-CONFIGURATIONAL SYMMETRIZED PROJECTOR QUANTUM MONTE CARLO METHOD 
FOR LOW-LYING EXCITED STATES OF THE HUBBARD MODEL}
\footnote{Contribution no. 1179 from the Solid State
and Structural Chemistry Unit } \\
\vspace{1cm}
{Bhargavi Srinivasan$^{3}$, S. Ramasesha$^{3,5}$ and H. R.
Krishnamurthy$^{4,5}$  }
{\footnote{e-mail: bhargavi@sscu.iisc.ernet.in,~ramasesh@sscu.iisc.ernet.in,\\
{~~~~~~hrkrish@physics.iisc.ernet.in}} }\\
\vspace{0.5cm}
{\it
$^{3}$Solid State and Structural Chemistry Unit  \\
\vspace{0.5cm}
$^{4}$Department of Physics  \\
Indian Institute of Science, Bangalore 560 012, India \\   
\vspace{0.5cm} 
$^{5}$Jawaharlal Nehru Centre for Advanced Scientific Research \\
Jakkur Campus, Bangalore 560 064, India \\
}
\end{center}
\vspace{1cm}

\pagebreak
\clearpage

\begin{center}
{\bf ABSTRACT}\\
\end{center}
We develop a novel multi-configurational Symmetrized Projector
Quantum Monte Carlo (MSPQMC) method to calculate excited state
properties of Hubbard models. We compare the MSPQMC results
for finite Hubbard chains with exact results (where available) or
with Density Matrix Renormalization Group (DMRG) results for longer 
chains.  The energies and correlation functions of the 
excited states are in good agreement with 
those obtained from the alternate techniques.

\section{Introduction}
In recent years,  numerical many-body techniques which have 
proved reliable for the study of the Hubbard
model are the Projector Quantum Monte Carlo (PQMC) 
method\cite{sorella,imada} and the 
Density Matrix Renormalization Group (DMRG)\cite{white} method 
(for 1-D and quasi 1-D systems).  These methods have
been predominantly ground state techniques. The usual procedure 
for obtaining excited states of a many-body
Hamiltonian in exact diagonalization schemes is to exploit the symmetries
of the system and block-diagonalize the Hamiltonian\cite{sraovb}. 
The lowest few
eigenvalues in each block can then be computed using standard numerical
procedures.  However, exact diagonalization schemes are limited
to rather small system sizes, unlike either the DMRG scheme or the PQMC
scheme. Implementation of symmetries in the latter schemes turns out to
be nontrivial although desirable. 

While the DMRG method could yield a 
few low-lying states, low-lying excited states of a 
chosen symmetry were inaccessible from this technique,
until its recent extension to incorporate 
crucial symmetries of a
system\cite{srsym}, which enables the method to target excited states 
as low-lying 
states in subspaces of a given irreducible 
representation of the symmetry group of the given system.

In contrast, the PQMC method has exclusively been a ground state technique
for fermionic systems.   Hitherto, it was not feasible to obtain
even the ground state of the 2-D Hubbard model for arbitrary filling 
because of the open-shell structure of the non-interacting ground
state\cite{bormann}. 
Here, we present a novel multi-configurational symmetrized
PQMC (MSPQMC) technique which makes it
possible, for the first time, to  obtain energies of {\it{excited states}}
of the Hubbard Hamiltonian. The technique is also applicable to the
ground state of open shell systems. We use the recently developed
symmetrized PQMC method\cite{pqmcc60} to improve the Monte Carlo 
estimates of properties in the targetted state. In the next section, the
formulation of the  MSPQMC method will be presented. In section 3, 
we demonstrate the method by applying it to Hubbard chains.
We also discuss numerical issues associated with the MSPQMC procedure. 

\section{The MSPQMC Method}

The single band  Hubbard Hamiltonian $\hat{H}$ for a system of
$N$ sites, may be written  as\cite{hub}, 
\begin{equation}
\hat{H} = \hat{H}_{0} + \hat{H}_{1} = -(\sum\limits_{\langle ij \rangle, \sigma} t_{ij}
\hat{a}^{\dagger}_{i\sigma} \hat{a}_{j\sigma} + h.c.)
+ U\sum_{i=1}^{N}\hat{n}_{i \uparrow}\hat{n}_{i \downarrow},
\label{hamiltonian}
\end{equation}
\noindent          
where the symbols have their usual meanings.
 
Using the projection ${\it ansatz}$, the lowest eigenstate,
$|\psi_{0}^{\Gamma}>$, in a 
given irreducible symmetry subspace $\Gamma$, of  
$\hat{H}$, can be projected from a trial wavefunction $|\phi^{\Gamma}>$  
as
\begin{equation}
|\psi_{0}^{\Gamma}> = \lim\limits_{\beta \rightarrow \infty} { {e^{- \beta{\hat H}}
|\phi^{\Gamma}>} \over {\sqrt{<\phi^{\Gamma}| e^{- 2 \beta{\hat H}} 
|\phi^{\Gamma}>}}},
\label{projec}
\end{equation}
\noindent
provided $|\phi^{\Gamma}>$ has a nonzero projection on to 
$|\psi_{0}^{\Gamma}>$. 
The trial wavefunction $|\phi^{\Gamma}>$ is usually formed 
from the molecular orbitals (MO) obtained
as eigenfunctions of the non-interacting part, $\hat{H}_0$, of the
full Hamiltonian. 
When the non-interacting ground state of a given system is a 
closed-shell state, the trial wavefunction
$| \phi^{\Gamma} >$ for obtaining the interacting ground
state can be chosen as a
single nondegenerate electronic configuration in the MO basis. However,
to obtain an excited state as the lowest eigenstate 
of the interacting 
model within a desired symmetry subspace from 
the projection ${\it ansatz}$,
it is usually necessary to choose $| \phi^{\Gamma} >$ to be
a specific linear combination of degenerate excited
MO-configurations as the trial wavefunction.
Such a linear combination can be obtained by operating with 
the group theoretic projection operator\cite{tinkham}
for the desired irreducible representation on a single
excited MO-configuration. In order to fix the total spin 
of the target state, we use the L\"owdin\cite{lowdin} projection 
operator to project out the desired spin state from the 
trial configuration transforming as $\Gamma$.
The projection procedure in eqn. (\ref{projec}) conserves 
the symmetry of the initial state and hence 
projects out the lowest energy excited state of the interacting 
model with that symmetry from the trial state. The trial 
state $|\phi^{\Gamma} >$ in general takes  the form,
\begin{equation}
|\phi^{\Gamma}> = \sum\limits_{j=1}^{p} c_j^{\Gamma} | \phi_{j}^{\Gamma} >~~;
~~| \phi_{j}^{\Gamma} > = | \phi_{j,\sigma}^{\Gamma} > 
| \phi_{j,-\sigma}^{\Gamma} >,
\label{multicon}
\end{equation}
where $p$ is the number of degenerate MO-configurations in the
symmetry adapted
starting wavefunction. An MO-configuration with $M_{\sigma}$ fermions 
of spin $\sigma$ in second
quantized form can be written as
\begin{equation}
|\phi_{j,\sigma}^{\Gamma}> = \prod\limits_{m=1}^{M_{\sigma}} 
\Bigl( \sum\limits_{i=1}^{N}
({\bf{\Phi}}_{\sigma}^{j\Gamma})_{im} \hat{a}_{i\sigma}^{\dagger} \Bigr) |0>
\label{twf}
\end{equation}
\noindent 
where ${\bf{\Phi}}_{\sigma}^{j\Gamma}$ is an $N \times M_{\sigma}$ 
sub-matrix of  the MO
coefficients whose row index, $i$, labels sites and the column
index,  $m$, labels the  MOs  occupied by  electrons of
spin $\sigma$, in the MO-configuration labelled by $j$ 
and $\Gamma$. The overlap of any two MO-configurations
expressed in this form is given by
\begin{equation}
<\phi_{j,\sigma}^{\Gamma}|\phi_{j^{\prime},\sigma}^{\Gamma}> =
det {\Bigl[} ({\bf{\Phi}}_{\sigma}^{j\Gamma})^{T}
({\bf{\Phi}}_{\sigma}^{j^{\prime}\Gamma}) {\Bigr]}.
\label{overlap}
\end{equation}

In the PQMC method for the Hubbard model, the projection 
operator $exp(-\beta \hat{H})$ is Trotter decomposed 
as $(exp(-\Delta \tau \hat{H}))^L$ with $L$ imaginary time slices of 
width $\Delta \tau$ $(\beta = L \times \Delta \tau)$.
This is followed by
a discrete Hubbard-Stratanovich (H-S) transformation\cite{temp} of the  
on-site interaction Hamiltonian at each site $i$ and each time-slice
$l$, in terms of Ising-like fields, $s_{il}$, leading to 
the result 
\begin{equation}
e^{-\Delta \tau \hat{H}} ~~=~~ 
\sum\limits_{\{{\bf{s}}_{l}\}}
{\hat{X}}_{\sigma}(l,{\bf{s}}_{l}){\hat{X}}_{-\sigma}(l,{\bf{s}}_{l})
\label{expdelta}
\end{equation}
\begin{equation}
{\hat{X}}_{\sigma}(l,{\bf{s}}_{l})=
exp[{{-\Delta \tau}\over {2}} \hat{H_0}] 
\sum\limits_{\{{\bf{s}}_{l}\}}
exp[\zeta_{\sigma} \lambda\sum\limits_{i} s_{il}\hat{n}_{i\sigma}
 - {\Delta\tau U \over 2} ]
exp[{{-\Delta \tau}\over {2}} \hat{H_0} ]
\label{xop}
\end{equation}
\noindent
where the summation is over all possible $N$-vectors
${\bf{s}}_{l}$ whose
$i^{th}$ components correspond to the H-S field,
$s_{il}$, $\zeta_{\sigma}$  is +1 (-1) for electrons with $\uparrow$ 
$(\downarrow)$ spin and the H-S parameter  $\lambda~=~$ $2 arctanh 
\sqrt{tanh(\Delta\tau U/4)}$.
Thus, 
\begin{equation}
e^{-\beta \hat{H}}~~=~~
\sum\limits_{\{s\}}
\hat{W}_{\sigma}(\{s\})\hat{W}_{-\sigma}(\{s\})~~=~~
\sum\limits_{\{s\}} \hat{W}(\{s\})
\label{expbeta}
\end{equation}
\begin{equation}
\hat{W}_{\sigma}(\{s\})~~=~~
{\hat{X}}_{\sigma}(L,{\bf{s}}_{L})\ldots
{\hat{X}}_{\sigma}(1,{\bf{s}}_{1} )  
\label{wop}
\end{equation}
\noindent
The action of each of the terms in the 
summation in eqn. (\ref{expdelta}) on a trial state of the form in eqn. (\ref{twf})
can be obtained as the left multiplication of the 
$N \times M_{\sigma}$ matrix $\bf{\Phi}^{j\Gamma}_{\sigma}$
by an $N \times N$ matrix,  ${\bf{B}}_{\sigma}(l,{\bf{s}}_{l})$,
given by
\begin{equation}
{\bf{B}}_{\sigma}(l,{\bf{s}}_{l}) = {\bf{b}}_{0}  {\bf{b}}_{1 \sigma}
(l,{\bf{s}}_{l})  {\bf{b}}_{0}
\label{bmatrix}
\end{equation}
\noindent
The matrix $ {\bf{b}}_{0}$ is given by $ exp [ - {\bf{K}}]$, with
$K_{ij} =   - {\Delta\tau\over2} t_{ij} $. 
The matrix  ${\bf{b}}_{1 \sigma}(l,{\bf{s}}_{l})$ is diagonal with
elements $ \delta_{ij} {1\over2} 
  exp [\zeta_{\sigma} \lambda s_{il} - {\Delta\tau \over 2} ]$.

To obtain the expectation value of an operator $\hat{O}$ 
in the targetted
state, when the trial state is a single MO-configuration,
 $< \hat{O} > = $ $(<\psi^{\Gamma}| \hat{O} |\psi^{\Gamma}>)/
 (<\psi^{\Gamma} | \psi^{\Gamma}>)$ we define states
$ |R(l,\{s_{R}\}) >$ and 
$< L(l,\{s_L\}) |$. The former
is obtained by projecting the trial wavefunction through the
right Ising lattice $\{s_R\}$ formed by time-slices 1 through $l$, while
the latter is obtained by projecting its transpose
through the left Ising lattice $\{s_L\}$ time-slices $L$ through 
$L-l+1$,
\begin{equation}
|\psi^{\Gamma} > \approx
\sum\limits_{\{s_{R}\}} |R^{\Gamma}(l,\{s_{R}\}) >
~~=~~ \sum\limits_{\{s_{R}\}} \hat{W}(\{s_R\}) |\phi^{\Gamma} >\\
\label{rket}
\end{equation}
\noindent
\begin{equation}
<\psi^{\Gamma}| \approx 
\sum\limits_{\{s_{L}\}} <L^{\Gamma}(l,\{s_{L}\}) |
~=~ \sum\limits_{\{s_{L}\}}<\phi^{\Gamma}| \hat{W}(\{s_L\})
\label{lket}
\end{equation}
\noindent
This allows us to express $<\hat{O}>$ as a weighted average,
$\sum\limits_{\{s\}}\omega^{\Gamma}(\{s\}) O^{\Gamma}(\{s\})$, with weights given by,
\begin{equation}
\omega^{\Gamma}(\{s\}) =
{{  <L^{\Gamma}(l,\{s_L\}) | R^{\Gamma}(l,\{s_R\}) > }
\over
{\sum\limits_{\{s\}}<\phi^{\Gamma} | \hat{W}(\{s\}) |\phi^{\Gamma} > }}
\label{omega}
\end{equation}
\begin{equation}
O^{\Gamma}(\{s\}) = 
{{  <L^{\Gamma}(l,\{s_L\}) | \hat{O} |R^{\Gamma}(l,\{s_R\}) > }
\over
{  <L^{\Gamma}(l,\{s_L\}) | R^{\Gamma}(l,\{s_R\}) > }}.
\label{oexpec}
\end{equation}
\noindent
For example, if the operator $\hat{O}$ is
the single-particle operator
$\hat{a}^{\dagger}_{k\sigma} \hat{a}_{l\sigma}$, $O^{\Gamma}(\{s\})$
takes the form,
\begin{equation}
O^{\Gamma}(\{s\}) ~~=~~
{\Bigl(} ( \bf{R}^{\Gamma}_{\sigma}(l,\{s\}) )
( \bf{L}^{\Gamma}_{\sigma}(l,\{s\}) \bf{R}^{\Gamma}_{\sigma}(l,\{s\} )^{-1}
( \bf{L}^{\Gamma}_{\sigma}(l,\{s\}) ) {\Bigr)}_{kl}.
\label{rlrinvl}
\end{equation}
\noindent
which is the $kl^{th}$ element of the
single-particle Green function,~~$G^{\Gamma}_{\sigma}(l,\{s\})$\cite{imada}.
If we weight average the property over all the Ising-configurations,
we would obtain the expectation value of that property in the 
targetted state, exact to within Trotter error. 
However, exhausting all Ising-configurations
in an averaging procedure is impractical and the denominator in
the eqn. (\ref{omega})  cannot be known explicitly. Therefore,
we resort to an importance sampling Monte Carlo (MC) estimation
in which a knowledge of the ratio of weights, $r$, for any two 
configurations $\{s^{\prime}\}$ and $\{s\}$,
$\omega^{\Gamma}(\{s^{\prime}\})/\omega^{\Gamma}(\{s\})$ is sufficient 
for obtaining property estimates. 
The ratio, $r$  is given by the ratio of inner products 
of the right and left projected states (using eqn. (\ref{overlap})) for the two
Ising-configurations,
\begin{equation}
r ~~=~~  \prod\limits_{\sigma} { {det
 {\Bigl(} {\bf{L}}^{\Gamma}_{\sigma}(l,\{s^{\prime}_{L}\})
{\bf{R}}^{\Gamma}_{\sigma}(l,\{s^{\prime}_{R}\}) {\Bigr)} }
\over
{det {\Bigl(} {\bf{L}}^{\Gamma}_{\sigma}(l,\{s_{L}\})
{\bf{R}}^{\Gamma}_{\sigma}(l,\{s_{R}\}) {\Bigr)} }  }
\label{rat}
\end{equation}
\noindent
Ising-configurations are 
generated by sequential single spin-flips through the lattice, 
examining each site at a given time slice, $l$, before 
proceeding to the next. 
This allows efficient computation of the ratio, $r$.
Using the heat bath algorithm, the new configuration
is accepted or rejected with a probability $r/(1+r)$.

The above procedure can be extended to a 
multi-configuration
trial function ($p>1$ in eqn.(\ref{multicon})) and the ratio of weights for 
Ising-configurations $\{s^{\prime}\}$ and $\{s\}$ takes the form,
\begin{equation}
{{\omega^{\Gamma}(\{s^{\prime}\})} \over  {\omega^{\Gamma}(\{s\})} }
~~=~~    
{{  <L^{\Gamma}(l,\{s^{\prime}\})|
R^{\Gamma}(l,\{s^{\prime}\}) > }
\over
 { <L^{\Gamma}(l,\{s\})|
R^{\Gamma}(l,\{s\}) > }}
\label{omegaprime}
\end{equation} 
\noindent
where the projected states $ |R^{\Gamma}(l,\{s_{R}\}) >$ 
and $<L^{\Gamma}(l,\{s_L\}) |$ are given by
\begin{equation}
|R(l,\{s_{R}\}) >~~=~~
\sum\limits_{j} c_{j}^{\Gamma}\prod\limits_{\sigma}
 |R_{\sigma}^{j\Gamma}(l,\{s_{R}\})>
 \label{rketmulti}
\end{equation}
\begin{equation}
<L(l,\{s_{L}\})| ~~=~~
\sum\limits_{j} c_{j}^{\Gamma}\prod\limits_{\sigma}
<L_{\sigma}^{j\Gamma}(l,\{s_{L}\}) |
\label{lketmulti}
\end{equation}
\noindent
with each state in the summations  obtained in a manner
analogous to the single-determinantal case. The ratio
(eqn. (\ref{omegaprime})) is now given as the ratio of sums of determinants
appearing in the numerator and denominator. Evaluating the
ratio hence turns out to be more time consuming than in the single
determinantal case.

Property estimates in the single-determinantal PQMC procedure,
with a sequential, single spin-flip algorithm, are carried out
by computing the Green function (eqn. (\ref{rlrinvl})) at the time slice at 
which a spin-flip is attempted.
This allows the use of an $O(N^2)$ updating 
algorithm for the Green function instead of the usual 
$O(N^3)$ direct algorithm. This is also applicable in
the MSPQMC method, except when
the states $ < L_{\sigma}^{j\Gamma}(l,\{s_L\}) |$ and 
$ | R_{\sigma}^{j\Gamma}(l,\{s_R\}) >$ are orthogonal. In this case 
we use
the explicit method of calculating matrix elements of the
single-particle Green function\cite{imada}.
The energies presented in this communication
have been obtained from the Green function (eqn. (\ref{rlrinvl})) 
estimated at the last time-slice, even when a spin flip is attempted 
at any intermediate time slice. Such an estimate of
energy will be more accurate, although it precludes the use
of an $O(N^2)$ updating algorithm. However, 
we still employ the single sequential spin-flip mechanism
as it reduces the number of  matrix multiplications 
involved in the computation of the Green function. 

In the MSPQMC method for excited states, we encounter the
negative sign problem even at half-filling although the number of
occurrences of the negative signs even at large $U/t$ is insignificant
(for $N=20, U/t=6$, the
fraction of the sample for which negative signs are encountered
is $\approx 10^{-4}$). The sign problem here arises because of
the phases with which the configurations in the trial state are
combined although products of individual determinants of 
up and down spin corresponding to 
$<L_{\sigma}^{j\Gamma}(l,\{s_{L}\}) | 
R_{\sigma}^{j^{\prime}\Gamma}(l,\{s_{R}\}) >$ are positive. The MSPQMC
method, besides being accurate for higher dimensional systems, also
has the advantage in one-dimensional systems over the DMRG method
in that the estimates of longer range correlations are as accurate 
as the nearest neighbour correlations.

One of the shortcomings of both single- and multi-configurational
PQMC calculations carried out as described above is that the
estimated properties do not reflect the
symmetries of the system, as the sampled Ising-configurations
do not have the full symmetry of the Hamiltonian. To obtain
symmetrized property estimates,
it is necessary to sample {\it all} symmetry related Ising-
configurations of every Ising configuration that is sampled. 
We have acheived this by a symmetrized PQMC (SPQMC) procedure
along the lines of the single-determinantal SPQMC method\cite{pqmcc60}. 
Such a symmetrized sampling reduces errors in estimates 
as the sample size is increased by a factor which is of the order of
the symmetry group, for a marginally small computational overhead.

\section{Results and Discussion}

The Hubbard chains and rings of $\le 14$ sites can be solved exactly
for energies of low-lying excitations in various symmetry subspaces. 
These results provide a strong check on the accuracy of the method. Besides, 
longer Hubbard chains can be solved to a high degree of accuracy 
by the DMRG method and we have compared our MSPQMC results 
against the DMRG results for longer chains ($N$ upto 20). 
We have computed the energies
of the lowest excited singlet state connected to the ground state by a dipole
transition as well the energies of the lowest triplet states for several
values of the Hubbard parameter $U$ and chain length $N$. 
The projection parameter $\beta$ is
set to 2.0 with a $\Delta \tau$ of 0.1, all in units of $t^{-1}$. 
All the estimates were carried out by averaging over 8000 spin flips
per Ising spin after allowing 2000 spin flips per Ising spin
for equilibration.  

In table $1$ we present the  MSPQMC energies for 
three different values of $U/t$. We note that the MSPQMC energies are
in very good agreement with exact results ($N \le 12$) or
high precision DMRG results for
($N > 12$). The  agreement is better at lower correlation strengths. 
While the DMRG energies are better than the MSPQMC energies\cite{srsym},
for the chosen states, it is worth noting that the MSPQMC 
excitation gaps have a better accuracy than the absolute energies as the
errors in the individual energies for the ground as well as the excited
states are comparable and have the same sign.

In fig. (1), we have shown the dependence of the two excitation 
gaps (with respect to the ground state) as a function of $U/t$ 
for different chain lengths. The 'optical' gap increases with
$U/t$ while decreasing with $N$ for a fixed $U/t$. The 'spin'
gap decreases with $U/t$ with the dependence on $N$ being similar
to the optical gap for a fixed $U/t$. This feature is in agreement
with exact as well as DMRG calculations. 

To conclude, we have shown that the PQMC method can be extended
to excited states using the symmetries of the Hamiltonian
via a multi- configurational formulation of the PQMC method.
Property estimates can be improved by symmetrized sampling
along the lines of a single-configuration SPQMC procedure.
To characterize excited states, we have also calculated 
other quantities such as bond-orders, spin correlations and 
charge correlations. All these quantities are in good agreement 
with exact/DMRG results and will be presented in a longer paper.\\
\noindent
{\bf{Acknowledgement:}} We thank Dr. Biswadeb Dutta for help with
the computer systems at JNCASR and Ms. Y. Anusooya and Mr. Swapan
Pati for help with exact and DMRG results.

\pagebreak
\begin{table}
\begin{center}
Table 1. MSPQMC energies (A) of lowest dipole allowed singlet and
lowest triplet excitations (in units of $t$) for
Hubbard chains of length 8 to 20 sites compared with
exact/DMRG results (B). For chain lengths ($\ge 14$ sites) 
comparison is with symmetrized DMRG results with a
cut-off of 150. Data for $U/t$=6.0 is with $\Delta\tau=0.05$. \\
\begin{tabular}{|c|c|cc|cc|} \hline 
\multicolumn{1}{|c|}{$U/t$} & \multicolumn{1}{c|}{$N$}&
\multicolumn{2}{c|}{Singlet}&
\multicolumn{2}{c|}{Triplet} \\ \hline 
      &    & A & B & A & B \\ \hline
1.0 &  6 &  -4.4766  & -4.4777&  -4.9331 &-4.9189\\
    &  8 &  -6.7881  & -6.7906&  -7.1378 &-7.1381\\
    & 10 &  -9.0185  & -9.0208&  -9.3066 &-9.3083\\
    & 12 &  -11.2030 & -11.2058 &  -11.4470 & -11.4516\\
    & 14 &  -13.3581 &  -13.3634&  -13.5767 &  -13.5785 \\
    & 16 &  -15.5007 &  -15.5032&  -15.6858 &  -15.6948  \\
    & 18 &  -17.6262 &  -17.6307&  -17.7970 &  -17.8037  \\
    & 20 &  -19.7429 &  -19.7494&  -19.8998 &  -19.9073  \\ \hline
4.0  &   6   & -0.3987 &  -0.4221&  -2.7088 &-2.6915\\
     &   8   & -1.8982 & -1.9301&  -3.9462  &-3.9165\\
     &   10  & -3.2633 &-3.3044 &  -5.0796  & -5.1151\\
     &   12  & -4.5378 & -4.6062&  -6.2310  & -6.2989\\
     &   14  & -5.7782 &   -5.8645&  -7.4085&  -7.4734\\
     &   16  & -7.0329 &  -7.0948 &  -8.5557&  -8.6418\\
     &   18  & -8.2260 &   -8.3059&  -9.6945 &  -9.8060\\
     &   20  & -9.3868 &   -9.5036&  -10.8457&  -10.9672 \\ \hline
6.0  &     6  & 1.9294  & 1.9212 &  -1.9793  &-1.9707\\
     &     8  & 0.7405  &0.6990 &  -2.9178   &-2.8677\\
     &     10 & -0.3477 & -0.3723&  -3.7362   &-3.7455\\
     &     12 & -1.3415 &  -1.3639&  -4.6146  & -4.6123\\
     &     14 & -2.2349 &  -2.3089 &  -5.4068  &  -5.4724\\
     &     16 & -3.1609 &  -3.2248 &  -6.2614  &  -6.3280\\
     &     18 & -4.0374 &   -4.1214&  -7.1203  &  -7.1805\\
     &     20 & -4.8692&   -5.0047&  -7.8726&  -8.0308 \\ \hline

\end{tabular}
\end{center}
\end{table}
\begin{center}
{\bf \underline{Figure Captions}} \\
\end{center}
\begin{description}
{\item {\bf Figure 1.}} "Optical" (filled symbols) and 
"Spin"(open symbols) gaps as a function of correlation
strength ($U/t$) for 
Hubbard chains of 16(squares), 18(circles) and 20(triangles) sites,
from the MSPQMC method.
\end{description}


\begin{thebibliography}{}
\bibitem{sorella} S. Sorella, E. Tosatti,  S. Baroni, R. Car and 
M. Parrinello, Int. J. Mod. Phys. {\bf B 1}, 993 (1988).
\bibitem{imada} M. Imada and Y. Hatsugai, J. Phys. Soc. Jpn. 
{\bf 58},  3752 (1989).
\bibitem{white} S. R. White, Phys. Rev. Lett. {\bf 69}, 2863 (1992);
Phys. Rev. {\bf B 48}, 10345 (1993).
\bibitem{sraovb} Z. G. Soos, S. Ramasesha, " Valence bond
theory and Chemical structure", ed. D. J. Klein and N. Trinajstic;
Elsevier: Amsterdam, 1990; p81. 
\bibitem{srsym} S. Ramasesha, Swapan K. Pati, H. R. Krishnamurthy,
Z. Shuai and J-L. Bredas, preprint no. cond-mat/9603162.
\bibitem{bormann} D. Bormann, T. Schneider and M. Frick,
Z. Phys. {\bf B 87}, 1 (1992).
\bibitem{pqmcc60} Bhargavi Srinivasan, S.Ramasesha and H. R. Krishnamurthy,
to appear in Phys. Rev. {\bf B} (1996).
\bibitem{hub} J. Hubbard,  Proc. Roy. Soc. Lond. {\bf 238}, A276 (1964);
{\it ibid.} {\bf 237}, A277 (1964); {\it ibid.}, {\bf 401}, A281  (1964). 
\bibitem{lowdin} P. O. L\"owdin, Adv. Phys. {\bf 5}, 40 (1956).
\bibitem{tinkham} M. Tinkham, "Group Theory and Quantum Mechanics",
McGraw-Hill Book Company, 1964.
\bibitem{temp} J. E. Hirsch, Phys. Rev. {\bf B 28}, 4059 (1983).
\end{thebibliography}
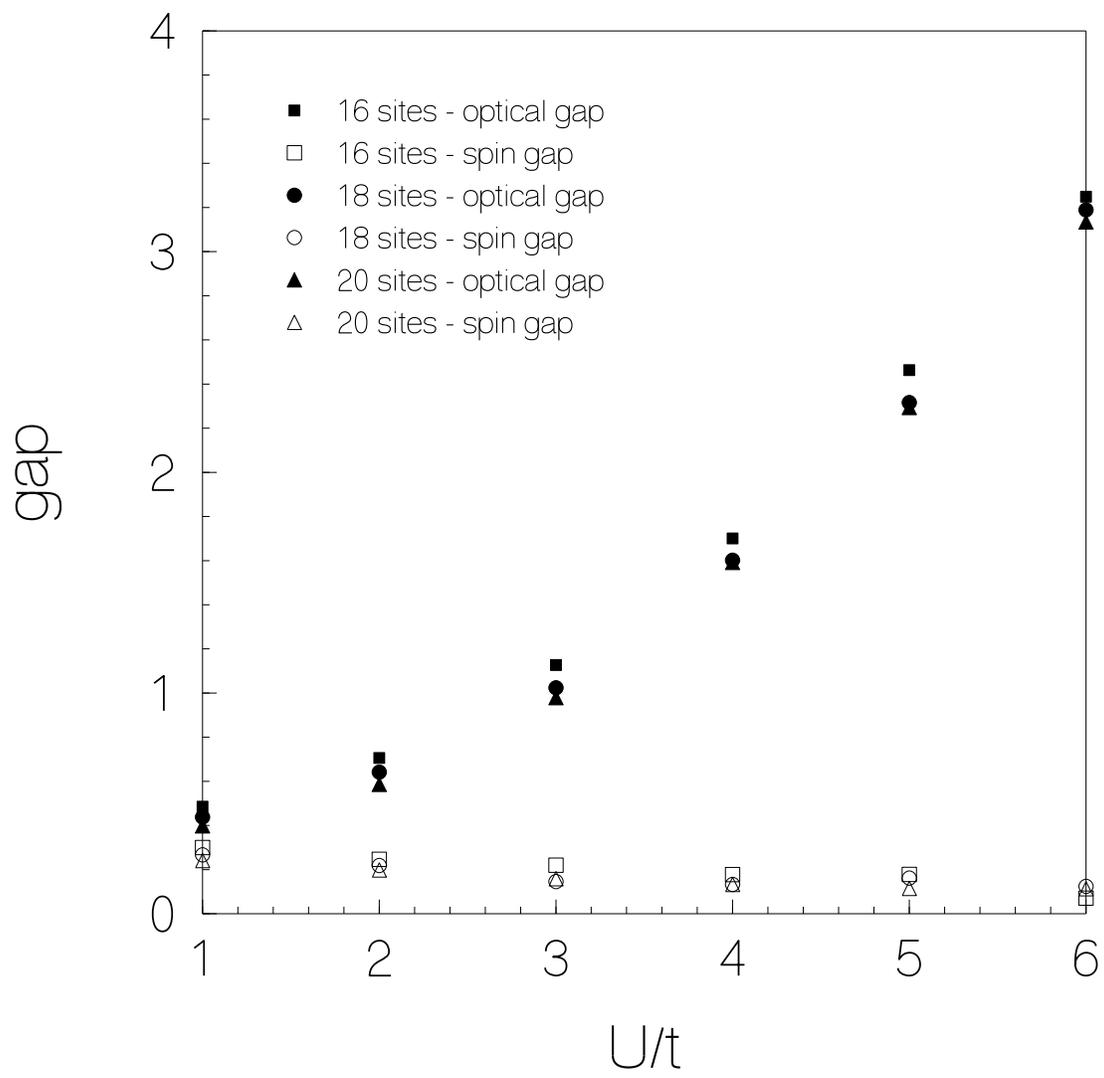
\end{document}